# Critical sizes for the submersion of alkali clusters into liquid helium

Casey Stark and Vitaly V. Kresin

*Department of Physics and Astronomy, University of Southern California,*
*Los Angeles, California 90089-0484*

**Abstract**

Alkali atoms do not stably embed in liquid helium-4 because the interatomic attractive potential is unable to overcome the short-range Pauli repulsion of the *s* electrons and the surface tension cost of the surrounding bubble. Similarly, small alkali complexes reside on the surface of helium nanodroplets instead of inside. However, as the size of the metal cluster increases, its van der Waals attraction to the helium matrix grows faster than the repulsive energies and above a certain size it should become favorable for clusters to submerge in the liquid. Based on an evaluation of the relevant energy terms, we characterize the bubble dimensions and estimate the critical submersion sizes. The latter range from $N\sim20$ for $Li_N$ and $Na_N$ to $N>100$ for $Rb_N$ in helium-4 and from $N\sim5$ for $Li_N$ and $Na_N$ to $N\sim20$ for $Cs_N$ in helium-3. These results are discussed in the context of nanodroplet pick-up experiments with alkali atoms.

# I. Introduction

A powerful technique for the study of cold atoms, molecules, and clusters is helium nanodroplet isolation (see, e.g., the reviews [1,2]), whereby a beam of superfluid $^4\text{He}_n$ nanodroplets picks up, entraps, and transports one or more atoms or molecules, while cooling them to sub-Kelvin temperatures. Interestingly, while most impurity atoms and molecules localize near the center of the droplet, alkali and alkaline atoms[3-5] prefer to remain bound in surface "dimples" because the Pauli repulsion between their $s$ valence electrons and those of the helium suppresses the interatomic attractive force.[6-8] This mirrors the notable weakness of the interaction between helium films and alkali substrates.[9,10]

As a result, one would expect alkali-metal clusters to have difficulty forming on the droplets: the addition of each new atom requires the dissipation of ~1 eV of condensation energy,[11,12] an amount that is so much greater than the strength of the alkali atom-helium surface binding (~2 meV[7]) that the cluster would be immediately blown off. Indeed, initially only dimers and trimers in predominantly high-spin configurations were identified on droplet surfaces,[13-15] consistent with the fact that such metastable states have a factor of ~10-50 lower binding energies. However, subsequent experiments have revealed that larger clusters do get produced by the He$_n$ pick-up technique: Na$_N$ and K$_N$ ions up to $N=25$ have been detected[16,17] (as well as Rb$^+_{1-5}$ and Cs$^+_{1-3}$). The probability of such large complexes assembling in spin-polarized configurations is exceedingly low,[18] therefore they are almost certainly in their ground-state metallic forms.

Do these larger clusters still hang onto the droplet surface, or do they sink into the liquid at some point? In this paper we present an estimate of the critical size for such a transition, which to our knowledge has not previously been addressed in any detail.

Similar to a submerged electron or alkali atom,[8] an alkali cluster within liquid helium will become surrounded by a bubble caused by the Pauli repulsion at the metal-helium interface (Fig. 1). Hence the total binding energy $\mathcal{E}_\text{solv}$ of the system will represent a balance of three interactions: the van der Waals attraction $E_{vdW}$ between the cluster and the helium medium, the short-range repulsion $E_{rep}$ of the alkali and helium atoms, and the surface tension cost of bubble formation. This needs to be compared with the energy $\mathcal{E}_\text{surf}$ of a cluster residing at the helium surface, where the surface tension and repulsion effects are diminished, but the van der Waals attraction is also reduced. Thus if at a particular cluster size $N_C$ the magnitude of $\mathcal{E}_\text{solv}$ exceeds that of $\mathcal{E}_\text{surf}$, solvation becomes beneficial for $N>N_C$.

Finite nanodroplet size may introduce a correction to the interaction energy, as can the diffuseness of the droplet surface.[19] For submerged alkali atoms, van der Waals interaction energies incorporating integration over a finite radius have been discussed in Ref. [20]. Since we will only consider relatively large droplets, such corrections can be neglected.

Thus we can formulate the problem as follows: how does the binding energy of a spherical metal cluster $M_N$ of $N$ atoms, total radius $R_N$, submerged into a large volume of helium and surrounded by a spherical cavity (bubble) of radius $R_b=R_N+\tau$ (Fig. 1),

$$\mathcal{E}_{solv} = E_{vdW}^{M_N-He} + E_{rep}^{M_N-He} + 4\pi\sigma R_b^2, \tag{1}$$

vary with *N,* and compare with the surface binding energy of the same cluster?

## II. Additive interaction approximation

This model is close to that considered by Ancilotto et al.[7] who examined the solvation of alkali atoms in helium by using the Lennard-Jones potential

$$V_{L-J}(r) = \frac{C_{12}}{r^{12}} - \frac{C_6}{r^6} \qquad (2)$$

with the coefficients for metal atom-helium atom interaction chosen to match Ref. [21] (see Table 1). The surface tension of the cavity walls included a curvature correction:

$$\sigma(R_b) = \sigma_0 \left(1 - \Delta/R_b\right) \qquad (3)$$

($\sigma_0 \approx 2.4 \times 10^{-7} E_h a_0^{-2}$, $\Delta \approx 0.64 a_0$, where $a_0$ is the Bohr radius and $E_h$ is the hartree energy unit[7]).

The most elementary way to extend this approach to a cluster of *N* alkali atoms ($R_N \approx a_0 r_s N^{1/3}$, $r_s$ is the Wigner-Seitz radius, see Table 1;) is to assume that the interaction (2) is additive, and to approximate both materials by uniform continua of atom number densities $n_M$ and $n_{He}$. Then the interaction energy is

$$E_{L-J}^{M_N-He} = \int_{0 \le r_{cl} \le R_N} dV_{cl} \int_{R_b \le r_{He} \le \infty} dV_{He} \cdot n_M n_{He} V_{L-J}\left(\left|\vec{r}_{cl} - \vec{r}_{He}\right|\right). \qquad (4)$$

Here $n_M \approx \left(\tfrac{4}{3}\pi r_s^3\right)^{-1} a_0^{-3}$ and[7] $n_{He} \approx 3.2 \times 10^{-3} a_0^{-3}$.

Integration over the cluster and helium volumes yields

$$\begin{aligned} E_{L-J}^{M_N-He} &= \frac{16\pi^2 n_{He} n_M}{135} C_{12} \frac{R_N^3 R_b^3 \left(5 R_N^4 + 14 R_N^2 R_b^2 + 5 R_b^4\right)}{\left(R_b^2 - R_N^2\right)^8} \\ &\quad - \frac{\pi^2 n_{He} n_M}{3} C_6 \left[\frac{2 R_N R_b \left(R_N^2 + R_b^2\right)}{\left(R_b^2 - R_N^2\right)^2} + \ln\left(\frac{R_b - R_N}{R_N + R_b}\right)\right], \end{aligned} \qquad (5)$$

The total binding energy is obtained by adding this expression to the surface tension energy in Eq. (1), and the equilibrium bubble radius is then determined by minimizing this total energy with respect to $R_b$. Carrying out this procedure, we make the useful and plausible finding that the metal-helium separation gap $\tau$ remains essentially constant for all cluster sizes.[22] These gap values (Table 1) are consistent with the bubble radii calculated for single atoms in Ref. [7]. With this knowledge of $R_b$, the total binding energy $\mathcal{E}_{\text{solv}}$ can now be calculated as a function of *N.*

## III. Van der Waals energies

We should keep in mind, however, that computing the van der Waals interaction between extended bodies by pairwise addition is well known to be inaccurate: one obtains answers that are of the correct order of magnitude, but potentially off by a significant numerical factor.[23] A more rigorous approach is based on evaluating the energy of the normal modes of the system, as defined by the boundaries and the dielectric functions of its constituent bodies. Fortunately, for the sphere-in-a-spherical-void geometry depicted in Fig. 1 the solution is available from Refs. [23,24]. In these references the answer is given for the interaction free energy at a finite temperature, as a summation over discrete (imaginary) Matsubara frequencies $i\xi_n$, where $\xi_n = n \cdot 2\pi k_B T$. Since our focus is on the low-temperature case, we can convert the summation over $n$ into an integration over $\xi_n$ via the Euler-Maclaurin formula, resulting in the van der Waals energy

$$E_{vdW}^{M_N-He} = \frac{\hbar}{2\pi} \sum_{m=1}^{\infty} (2m+1) \int_0^{\infty} d\xi \ln \left[ 1 - \frac{m(m+1)(\varepsilon_M - 1)(\varepsilon_{He} - 1)}{[m\varepsilon_M + m + 1][(m+1)\varepsilon_{He} + m]} \left( \frac{R_N}{R_b} \right)^{2m+1} \right]. \quad (6)$$

The gap between the metal particle and the bubble wall is assigned $\varepsilon=1$, and the metal ($\varepsilon_M$) and helium ($\varepsilon_{He}$) dielectric functions are evaluated at the imaginary frequencies $i\xi$.

For the dilute and weakly interacting liquid helium we can relate $\varepsilon_{He}$ to the atomic dynamical polarizability, approximating the characteristic resonance frequency with the atomic ionization potential ($I_{He}=0.90\ E_h$):[23]

$$\begin{aligned}\varepsilon_{He} &\approx 1 + 4\pi n_{He} \alpha_{He}(i\xi) \\ &\approx 1 + 4\pi n_{He} \frac{e^2}{m_e} \frac{2}{I_{He}^2 + \xi^2}.\end{aligned} \quad (7)$$

Here $m_e$ is the electron mass and 2 is the number of valence electrons in the atom, i.e., the oscillator strength.

For the alkali metals, in turn, we use the plasma-pole approximation:

$$\varepsilon_M = 1 + \frac{\tilde{\omega}_p^2}{\xi^2}, \quad (8)$$

where $\tilde{\omega}_p$ is the metal plasma frequency reduced by the so-called electron spillout correction. The correction reflects the fact that in a small nanocluster a non-negligible fraction of the electron cloud extends beyond the boundary of the ionic core, decreasing the effective electron density and thereby the plasma frequency. A convenient way to parametrize it is by introducing an adjusted electron cloud radius[25] $R=R_N+\delta$, i.e., by writing

$$\tilde{\omega}_p^2 \approx \frac{4\pi n_M e^2}{m_e} \frac{R_N^3}{(R_N + \delta)^3} \quad (9)$$

We use the value $\delta=2a_0$ which approximately fits the response properties (surface plasmon resonances and electric polarizabilities) of alkali clusters.[26,27] As will be seen later, the spillout has a strong effect on the submersion size.

Fig. 2 illustrates that as expected, pairwise addition of interatomic $C_6/r^6$ van der Waals attraction [second term in Eq. (5)] and the many-body expression (6) yield values that are analogous, but numerically distinct. Since the critical submersion sizes $N_c$ are determined by the balance of several relatively shallow curves, this distinction leads to strong deviations in $N_c$.

We now employ Eqs. (6)-(9) as the attraction term $E_{vdW}^{M_N-He}$ in the total energy expression

(1). For the repulsion term $E_{rep}^{M_N-He}$, we continue to use the original $C_{12}$ term on the right-hand side of Eq. (5), since its near-range character makes it quite a bit more local than the van der Waals attractive interaction.

Using this composition, the total binding energy $\mathcal{E}_{solv}$ from Eq. (1) can again be minimized to find the optimal bubble radius $R_b$ (that is, the optimal metal-helium separation gap $\tau$) for a given cluster size $R_N$. The resultant values of $\tau$ are listed in Table 1. Again, they are found to be independent of cluster size,[22] and in fact very close to the values obtained from the more elementary calculation in Sec. II.

Using thus determined bubble dimensions, the total binding energy $\mathcal{E}_{solv}$ can now be calculated as a function of $N$. Fig. 3 shows an example of the behavior of the three contributions entering Eq. (1) and of the total binding energy. This quantity can now be compared with the binding energy of the competing geometry: a nanocluster residing at the helium surface.

### IV. Surface binding

For an estimate of $\mathcal{E}_{surf}$, the binding energy of a nanocluster residing at the surface of a helium droplet or a helium reservoir, we adopt the simple picture of a metal particle positioned a short distance $d$ away from a plane surface of liquid helium. This evaluation does not incorporate the appearance of a size- and material-dependent dimple structure in the reduced-density surface layer of the liquid,[1-3,28] but it has the advantage of being analytically tractable while yielding results of a reasonable magnitude. This picture involves two terms:

$$\mathcal{E}_{surf} = E_{vdW,surf}^{M_N-He} + E_{rep,surf}^{M_N-He}. \tag{10}$$

The first is the van der Waals attraction between a metallic sphere and a dielectric plane, and is given[23] by an analogue of Eq. (6):

$$E_{vdW,surf}^{M_N-He} = -\frac{\hbar}{8\pi}\left(\frac{R_N}{d} + \frac{R_N}{2R_N+d} + \ln\frac{d}{2R_N+d}\right)\int_0^\infty d\xi \frac{(\varepsilon_M-1)(\varepsilon_{He}-1)}{(\varepsilon_M+1)(\varepsilon_{He}+1)}. \tag{11}$$

For the minimum cluster-liquid separation distance we take the calculated[28] width of the $^4$He surface region: $d \approx 5$ Å. The dielectric functions are evaluated as in Eqs. (7)-(9).

The second term in Eq. (10) is calculated by pairwise summation of the Lennard-Jones $C_{12}$ repulsion interaction over the metallic nanocluster sphere and the bulk liquid. An expression for this sum has been derived in Ref. [29]:

$$E_{rep,surf}^{M_N-He} = \frac{\pi^2 n_{He} n_M}{7560} C_{12}\left[\frac{8R_N+d}{(2R_N+d)^7} + \frac{6R_N-d}{d^7}\right]. \tag{12}$$

As one might anticipate, this term makes only a small contribution to the total energy.

## V. Critical sizes for submersion

Fig. 3 illustrates the evolution of $\mathcal{E}_{\text{solv}}$ and $\mathcal{E}_{\text{surf}}$. Whereas the cluster-surface interaction, $\mathcal{E}_{\text{surf}}$, is attractive throughout, in the case of submersion the repulsive parts of the total energy initially dominate. As the cluster size increases, $\mathcal{E}_{\text{solv}}$ first changes sign and then finally exceeds the magnitude of $\mathcal{E}_{\text{surf}}$ at a certain cluster size $N_c$. This signifies that particles larger than this find it energetically beneficial to become submerged within the liquid. The corresponding critical cluster sizes $N_c$ are tabulated in Table 1.

The very large critical size calculated for Cs shows that for this metal the ratio of van der Waals attraction to short-range repulsion is particularly weak. This ties in with the experimental fact that bulk Cs is the only material not wetted by superfluid helium.[30] In fact, in view of the model's approximate nature, the value of $N_c \sim 600$ should be taken not literally, but rather as a signal of failing to submerge.[31]

Although the electron spill-out correction to $\omega_p$ may seem moderate, leaving it out would decrease all $N_c$ values by approximately 60% to $N_c \approx 10, 10, 30, 55, 200$ for Li, Na, K, Rb, Cs, respectively. The use of pairwise-additive $C_6$ energy from Sec. II would lead to $N_c \approx 5, 5, 10, 15, 25$. This estimate appears too low (for example, it would imply too efficient a wetting for Cs; furthermore, ionization profiles of sodium clusters on $^4$He nanodroplets suggest[16] that Na$_n$ remains on the surface at least up to $n \sim 10$) and illustrates the merit of treating van der Waals interactions in a many-body fashion.[23]

An analogous model calculation can be carried out for $^3$He whose atomic number density and surface tension are approximately 25% and 60% lower than those of $^4$He, the latter factor reducing the energy cost of creating a cavity. Using the same values for $C_6$, $C_{12}$, $\Delta$ as before, and additional parameters from Refs. [7,28] ($n_{He} \approx 2.4 \times 10^{-3} a_0^{-3}$, $\sigma_0 \approx 1.0 \times 10^{-7} E_h a_0^{-2}$, $d \approx 6$ Å), we obtain numbers that are also listed in Table 1. They are substantially smaller than in the case of $^4$He.

While the evaluation of $N_c$ presented here obviously is to be viewed as an estimate that involves a number of approximations (e.g., inexact treatment of the helium surface tension and metal electron spill-out, pairwise addition of repulsive forces, a fairly schematic model of surface binding), it offers useful guidance regarding the balance of the relevant energies, and regarding the relative affinity of various alkali nanoclusters for submersion in liquid helium and helium droplets. For example, the fact that large Na and K but only small Rb and Cs clusters ions were detected upon pick-up and photoionization in Ref. [17] correlates with the higher binding energies of the former two elements and therefore an enhanced stability against desorption. In experimental work on nanodroplet pickup, it may be possible to establish the transition from surface to interior location of alkali clusters from the threshold shape of the electron-impact ionization yield: this shape reflects the relative contributions of Penning (impurity interaction with He*) vs. charge-transfer (interaction with He$^+$) ionization channels, the former making a larger contribution for surface-bound species.[5,16]

We also hope that this work will motivate further theoretical investigations at a microscopic level. It is in agreement with theory and experiment that alkali atoms on both $^4$He and $^3$He prefer surface states,[7,28,32] as do alkali dimers and trimers on $^4$He (as described in Sec. I), but we are not aware of detailed theoretical studies for larger clusters.

In summary, we have considered solvation of alkali nanoclusters in liquid helium by analyzing the competition between the metal-helium van der Waals attraction on one hand, and the costs of the short-range repulsion and the surface tension of the surrounding bubble on the other hand. The bubble dimensions and the binding energy were evaluated using first the model of pairwise addition of the Lennard-Jones potential [Eq. (2)] and then a more accurate many-body expression for the van der Waals interaction [Eq. (6)]. The bubble gap separating the metal and the helium has been found to remain essentially constant for all cluster sizes. The critical cluster sizes at which full submersion becomes energetically favorable have been estimated. They increase with the atomic mass and are substantially smaller for $^3$He than for $^4$He.

We thank the referees for their constructive suggestions. This work was supported by the National Science Foundation and by the USC Undergraduate Research Associates Program.

**Table 1.** Density and interatomic interaction parameters of the alkalis, and the calculated bubble dimensions and critical submersion cluster sizes for liquid $^4$He and $^3$He.

| Cluster material | $r_s$ [a] | $C_6$ ($E_h a_0^6$) [b] | $C_{12}$ ($E_h a_0^{12}$) [b] | $^4$He | | $^3$He | |
|---|---|---|---|---|---|---|---|
| | | | | $\tau$ ($a_0$) [c] | Submersion cluster size $N_c$ [d] | $\tau$ ($a_0$) [c] | Submersion cluster size $N_c$ [d] |
| Li | 3.25 | 22.5 | $2.08 \times 10^7$ | 8.0 | 23 | 8.3 | 8 |
| Na | 3.93 | 24.7 | $2.78 \times 10^7$ | 7.8 | 21 | 8.1 | 7 |
| K | 4.86 | 38.9 | $8.47 \times 10^7$ | 8.6 | 78 | 8.9 | 13 |
| Rb | 5.20 | 44.6 | $1.11 \times 10^8$ | 8.8 | 131 | 9.1 | 16 |
| Cs | 5.63 | 51.2 | $1.71 \times 10^8$ | 9.1 | 625[e] | 9.4 | 24 |

(a) Ref. [33].

(b) Ref. [7].

(c) Metal-helium gap calculated by using Eq. (6). Using the additive $C_6$ term in Eq. (1) or neglecting the spillout effect reduces $\tau$ by only 4%-7%.

(d) Sizes calculated using the expressions in Sec. IV and V.

(e) Indicative of lack of submersion, see text.

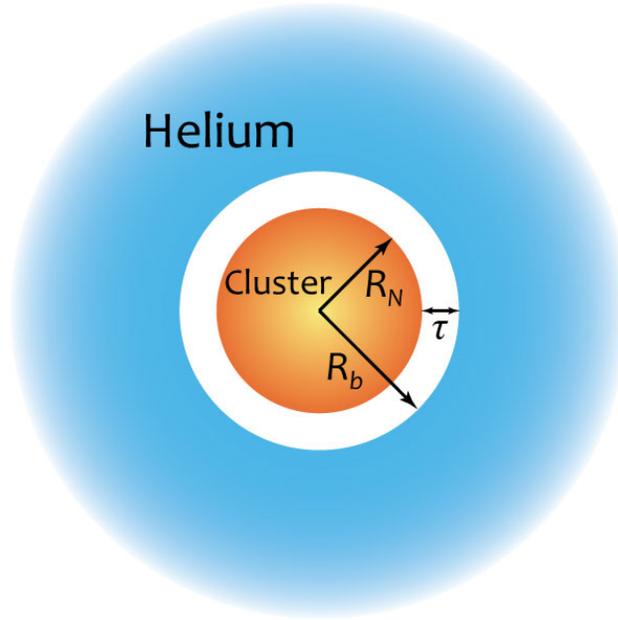

**Fig. 1.** Model of an alkali nanocluster submerged in a liquid helium bath, residing within a bubble.

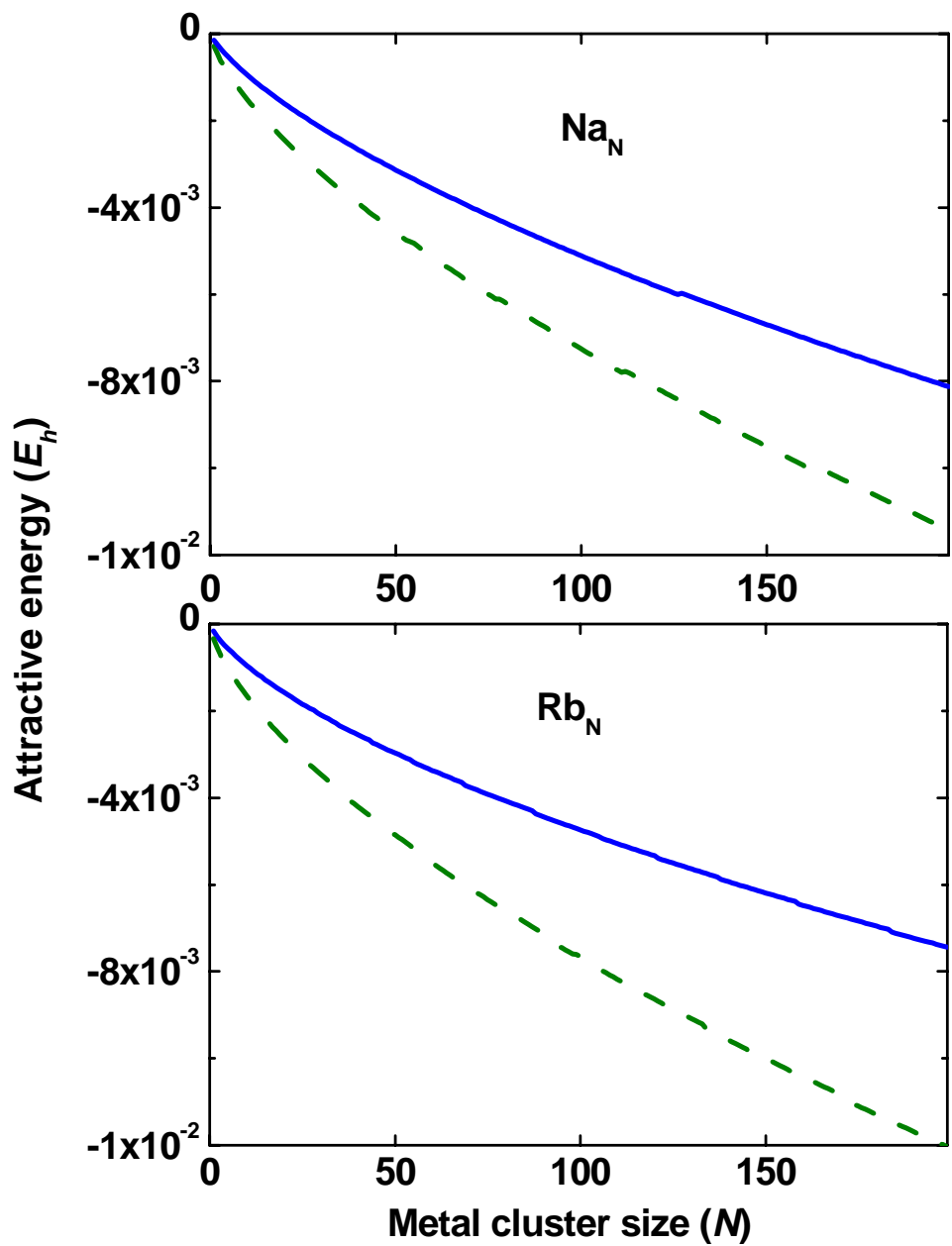

**Fig. 2.** Examples of differences in the attractive cluster-helium van der Waals energy calculated by pairwise addition of the interatomic $C_6/r^6$ potential (green dashed line) and the full expression in Eq. (6) (blue solid line). The energy scale is in atomic units, and $N$ is the number of atoms in the cluster particle. Top plot: sodium, bottom plot: rubidium.

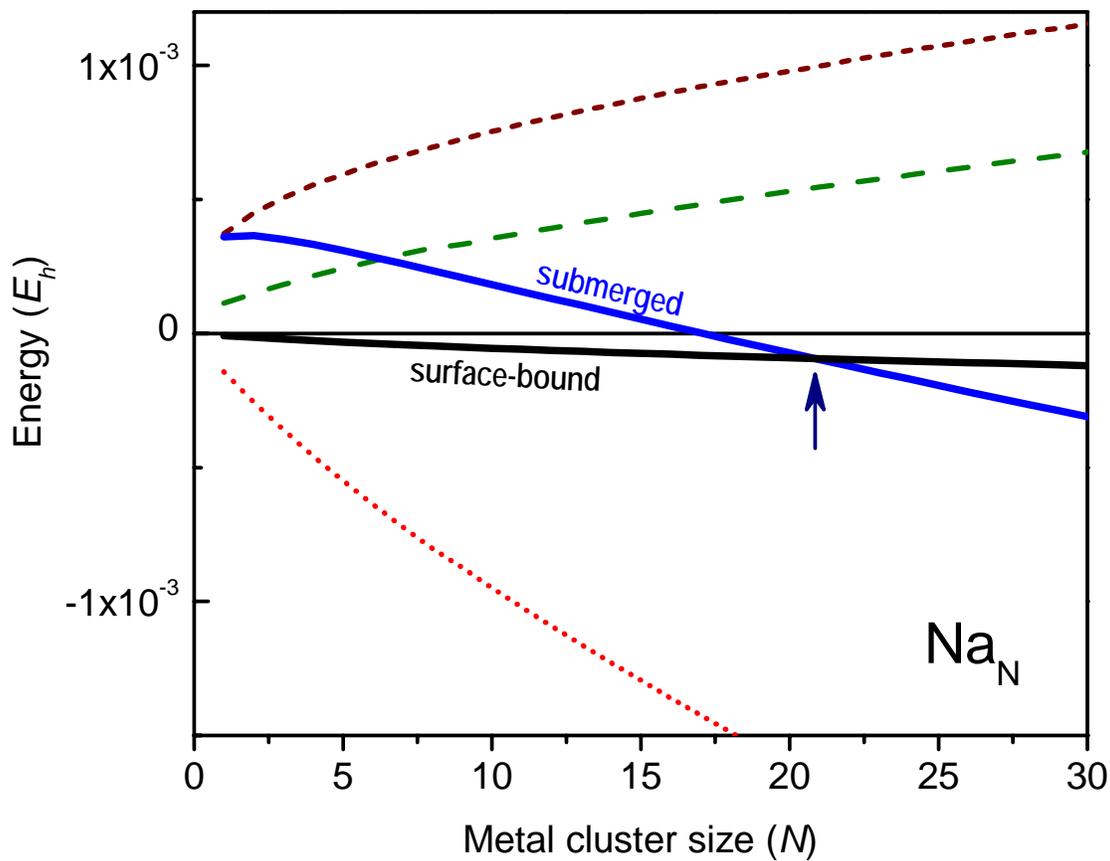

**Fig. 3.** Appearance of cluster submersion with increasing size, for the example of Na$_N$. The three dashed curves depict the three contributions to Eq. (1): van der Waals attraction [Eqs. (6)-(9), red dots], short-range repulsion [first term in Eq. (5), green long dash], and the surface tension energy of the cavity wall [last term in Eq. (1) with the use of Eq. (3), brown short dash]. The solid blue line is the sum of these, i.e., the total binding energy of a solvated cluster, $\mathcal{E}_{solv}$. The solid black line is the binding energy at the surface, $\mathcal{E}_{surf}$, Eqs. (10)-(12). Submersion corresponds to the crossing of these lines, as marked by an arrow.